\documentclass{PoS}
\usepackage{capt-of}

\title{Canonical approach to the finite density QCD with winding number expansion}

\ShortTitle{Canonical approach to the finite density QCD with winding number expansion}

\author{Atsushi Nakamura
\\
RIISE, Hiroshima University
\\
Higashi-Hiroshima 739-8521, Japan
\\
        E-mail: \email{nakamura@riise.hiroshima-u.ac.jp}}

\author{Shotaro Oka
\\
        Institute of Theoretical Physics, Department of Physics,
 	Rikkyo University
\\
Toshima-ku, Tokyo 171-8501, Japan
\\
        E-mail: \email{okasho-hato@rikkyo.ac.jp}}

\author{\speaker{Yusuke Taniguchi}%
\\
 Graduate School of Pure and Applied Sciences,
      University of Tsukuba \\
Tsukuba, Ibaraki 305-8571, Japan\\
       E-mail: \email{tanigchi@het.ph.tsukuba.ac.jp}}

\abstract{
The canonical partition function is related to the grand canonical one through 
the fugacity expansion and is known to have no sign problem.
In this paper we perform the fugacity expansion by a method of the hopping parameter expansion
in temporal direction for the lattice QCD: winding number expansion.
The canonical partition function is constructed for $N_f=2$ QCD starting from 
gauge configurations at zero chemical potential.
After derivation of the canonical partition function we 
calculate hadronic observables like chiral condensate and quark number
density and the pressure at the real chemical potential.
}

\FullConference{The 32nd International Symposium on Lattice Field Theory\\
		 23-28 June, 2014\\
		 Columbia University New York, NY}

\begin{document}

\section{Introduction}

The grand canonical ensemble is a difficult subject to treat in lattice QCD because of the sign problem.
The canonical partition function is related to the grand canonical one through 
the fugacity expansion and is known to have no sign problem as will be briefly reviewed in Sec.~2.
In this paper we perform the fugacity expansion by a method of the hopping parameter expansion as is
discussed in Sec.~3.
The canonical partition function is evaluated for $N_f=2$ QCD starting from 
gauge configurations at zero chemical potential in Sec.~5.
After derivation of the canonical partition function we study the
chemical potential dependences of hadronic observables like chiral
condensate and quark number density in Sec.~6.

\section{Canonical partition function}

It is a well known fact that the grand canonical ensemble and the canonical one are equivalent each other.
This is shown by a simple equation to relate the grand canonical partition function $Z_G(\mu, T, V)$ and
the canonical $Z_C(n, T, V)$
\begin{eqnarray}
Z_G(\mu, T, V)=\sum_{n=-\infty}^\infty Z_C(n,T,V)\xi^n,
\quad
\xi=e^{\mu/T}.
\label{ZG}
\end{eqnarray}
This is a so called fugacity expansion formula with the fugacity $\xi$.
The inverse of this expansion is given by using the Cauchy's integral theorem
\begin{eqnarray}
Z_C\left(n,T,V\right)=\oint\frac{d \xi}{2\pi i}\xi^{-n-1}Z_G(\xi,T,V),
\end{eqnarray}
where we assume that the partition function $Z_G(\xi,T,V)$ has singularities only at $\xi=0, \infty$
corresponding to trivial points $\mu/T=\pm\infty$ and adopt an appropriate contour around the origin.
Here we notice that a phase transition point $\xi_c$ is not a singularity of the partition function but it is
rather a zero of $Z_G(\xi,T,V)$ in $V\to\infty$ limit (Lee-Yang zeros) and does not affect the Cauchy's integral.

Now it is free to change the contour to a unit circle $\xi=e^{i\theta}$ and the contour integral turns out
to be a Fourier transformation \cite{Hasenfratz:1991ax}
\begin{eqnarray}
Z_C\left(n,T,V\right)=\int_0^{2\pi}\frac{d\theta}{2\pi}e^{-in\theta}Z_G(e^{i\theta},T,V).
\label{Fourier}
\end{eqnarray}
A standard Monte Carlo simulation is possible for the lattice QCD since the chemical potential is set to be
pure imaginary $\mu/T=i\theta$ and the Boltzmann weight is real positive for two flavors.

However a difficulty of the sign problem is preserved unfortunately since there remains a frequent
cancellation in plus and minus sign in a numerical Fourier transformation.
In order to get the canonical partition function accurately for large quark number $n$ we need a very fine
sampling of $Z_G(e^{i\theta},T,V)$ in $\theta$.
This requires a heavy computational cost because we need to evaluate the Dirac operator determinant
for the QCD grand partition function.
In this paper we shall solve this problem by a direct expansion of the Dirac determinant in terms
of the fugacity.

\section{Winding number expansion}
\label{sec:WNE}

We consider the lattice QCD grand partition function given in the path integral form
\begin{eqnarray}
Z_G(\xi,T,V)=\int DU {\rm Det}D_W(\xi;U)e^{-S_G(U)},
\end{eqnarray}
where we adopt the Wilson Dirac operator
\begin{eqnarray}
&&
D_W=1-\kappa Q,\quad
Q=\sum_{i=1}^3\left(Q_i^++Q_i^-\right)+e^{\mu a}Q_4^++e^{-\mu a}Q_4^-,
\\&&
\left(Q_\mu^+\right)_{nm}=\left(1-\gamma_\mu\right)U_\mu(n)\delta_{m,n+\hat{\mu}},\quad
\left(Q_\mu^-\right)_{nm}=\left(1+\gamma_\mu\right)U_\mu^\dagger(m)\delta_{m,n-\hat{\mu}}.
\end{eqnarray}
Both the chemical potential $e^{\pm\mu a}$ and the hopping parameter
$\kappa$ appear in front of the temporal hopping term simultaneously.
The fugacity expansion of the determinant shall be given by using the
hopping parameter expansion.

Instead of the determinant we consider the hopping parameter expansion of
\begin{eqnarray}
{\rm Tr} {\rm Log} D_W(\mu)={\rm Tr} {\rm Log}\left(I-\kappa Q\right)
=-\sum_{n=1}^\infty\frac{\kappa^n}{n} {\rm Tr}(Q^n).
\label{hpe}
\end{eqnarray}
Since every term has a trace the quark hoppings should make a loop.
A non-trivial chemical potential dependence appears when the quark
hoppings make a loop in the temporal direction.
If one of the term has a $n$ times winding loop in positive temporal direction the chemical potential
dependence is $e^{\mu a nN_t}$ where $N_t$ is a temporal lattice length and this is nothing but
$e^{n\mu/T}=\xi^n$.
Counting the winding number in temporal direction for each term in (\ref{hpe}) we get the winding number
expansion of the TrLog
\begin{eqnarray}
{\rm Tr}{\rm Log}D_W(\mu)=\sum_{n=-\infty}^\infty w_n\xi^n
\end{eqnarray}
Regrouping the summation we have a fugacity expansion of the determinant
\cite{Li:2010qf}
\begin{eqnarray}
{\rm Det}D_W(\xi;U)=\exp\left(\sum_{n=-\infty}^\infty w_n\xi^n\right)
=\sum_{n=-\infty}^\infty z_n(U)\xi^n.
\label{regroup}
\end{eqnarray}
In this approach the fugacity dependence of the determinant or the partition function is given analytically.
A numerical Fourier transformation (\ref{Fourier}) can be executed safely at any high precision we want.
In this paper the regrouping (\ref{regroup}) is done by the Fourier transformation.

The winding number expansion is done for gauge configurations generated at $\mu=0$ or purely imaginary
value.
This procedure corresponds to the standard reweighting method
\begin{eqnarray}
Z_G(\xi,T,V)=\int DU \frac{{\rm Det}D_W(\mu)}{{\rm Det}D_W(\mu_0)}
{\rm Det}D_W(\mu_0)e^{-S_G}.
\end{eqnarray}

\section{Numerical setup}

\begin{figure}
\begin{minipage}[t]{.55\textwidth}
\centering
\vspace{0pt}
\begin{tabular}{|c|c|c|c|}
\hline
$\beta$ & $\kappa$ & PCAC mass & $m_\pi/m_\rho$ \\
\hline
$0.9$ & $0.137$ & $0.17(13)$ & $0.8978(55)$\\
$1.1$ & $0.133$ & $0.18(19)$ & $0.9038(56)$\\
$1.3$ & $0.138$ & NA & $0.809(12)$\\
$1.5$ & $0.136$ & NA & $0.756(13)$\\
$1.7$ & $0.129$ & $0.168(21)$ & $0.770(13)$\\
$1.9$ & $0.125$ & $0.1076(68)$ & $0.714(15)$\\
$2.1$ & $0.122$ & $0.1259(11)$ & $0.836(47)$\\
\hline
\end{tabular}
\captionof{table}{$(\beta,\kappa)$ for the numerical simulation.
The PCAC mass and $m_\pi/m_\rho$ are also given.}
\label{table}
\end{minipage}
\hfill
\centering
\begin{minipage}[t]{.4\textwidth}
\centering
\vspace{0pt}
\includegraphics[width=0.96\textwidth]{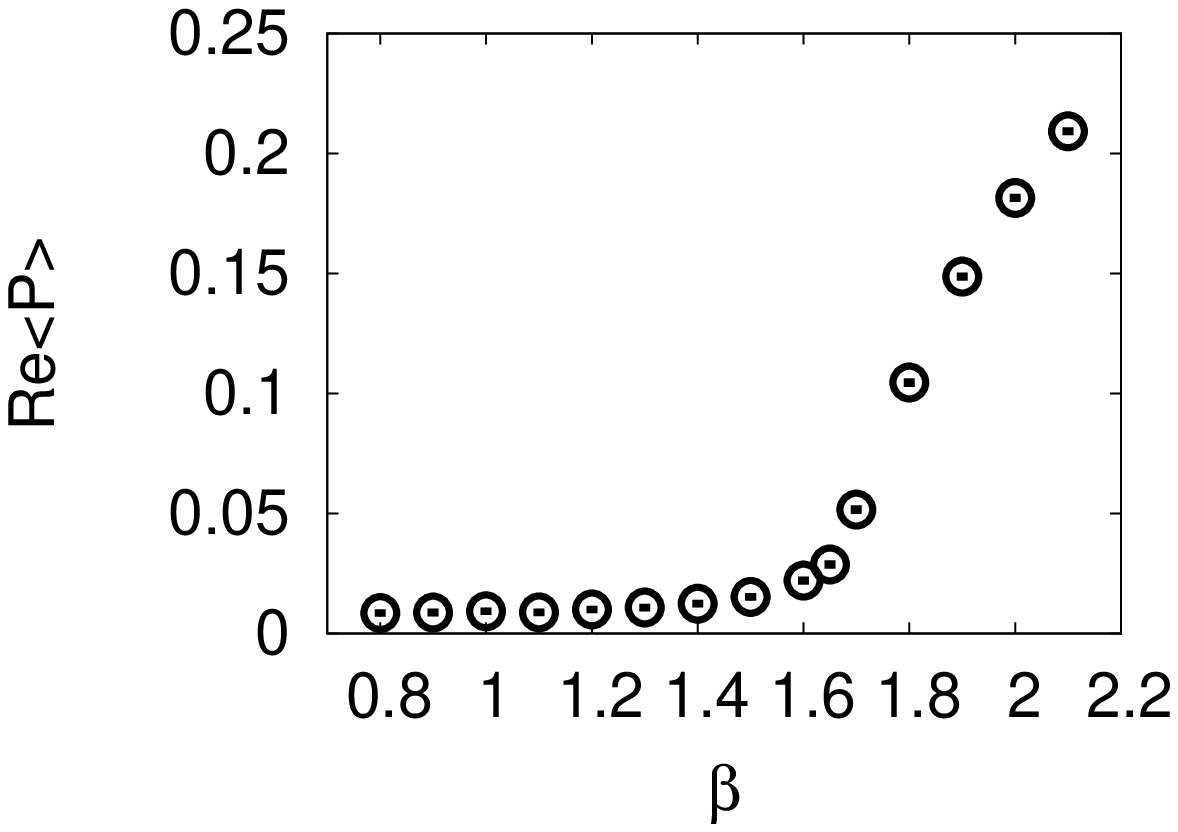}
  \caption{The Polyakov loop as a function of $\beta$.}
  \label{fig:polyakov}
\end{minipage}
\end{figure}

We adopt the Iwasaki gauge action and the improved Wilson fermion action
with the clover term.
The number of flavors is set to $N_f=2$ with degenerate masses.
The APE stout smeared gauge link is used for those in the fermion
action including the clover term.
The number of smearing is four and the parameter is set to $\rho=0.1$.
The clover coefficient is fixed to $c_{\rm SW}=1.1$ \cite{Taniguchi:2013cxa}.
We adopt $8^3\times4$ lattice.
The numerical parameters $(\beta,\kappa)$ are given in table \ref{table}.
A wide range of $\beta$ is covered from high temperature $\beta=2.1$ to low temperature $0.9$.
It seems to be that both the confining and deconfining region are well
covered, which is inferred from a behavior of the Polyakov loop in
Fig.~\ref{fig:polyakov}.
The hopping parameter is selected in order that the hopping parameter expansion works well, for which
$m_\pi/m_\rho$ turns out to be $0.7$ - $0.9$.
Maximal order of the hopping parameter expansion is set to $480$ so that max winding number in temporal
direction is $120$.

\section{Canonical and grand canonical partition function}

\begin{figure}
 \begin{center}
  \includegraphics[width=5.7cm]{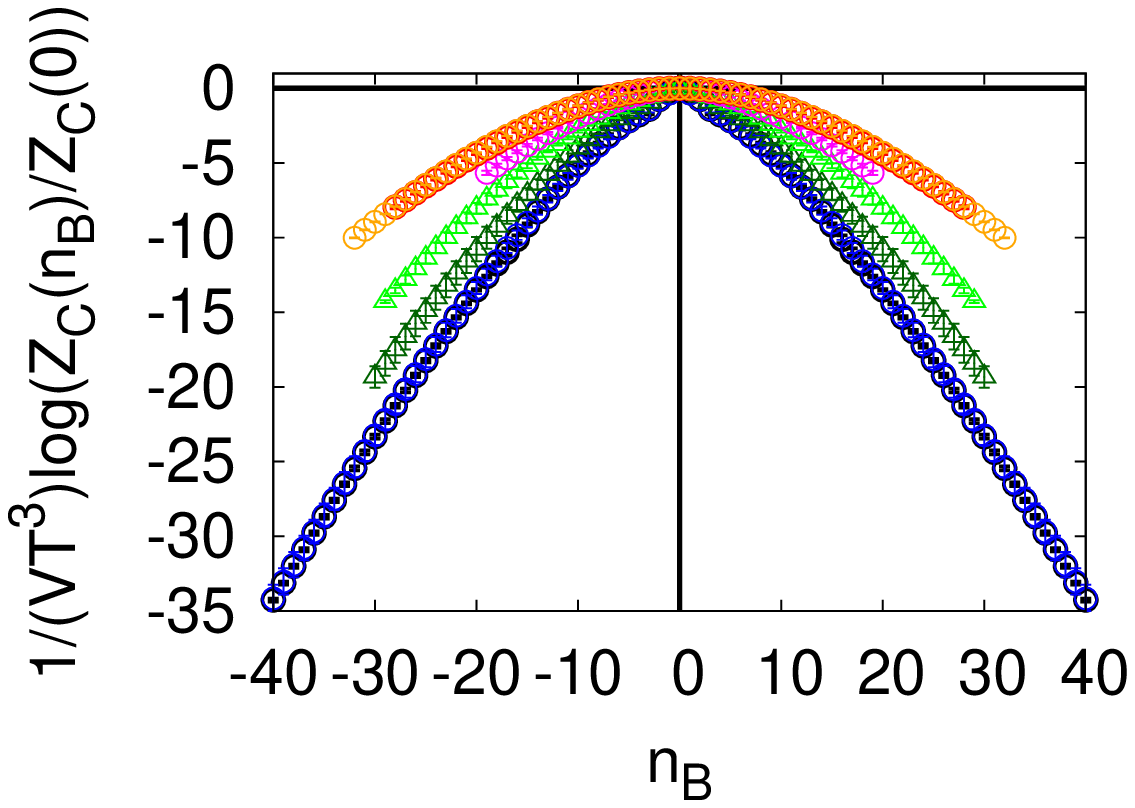}
  \includegraphics[width=5.7cm]{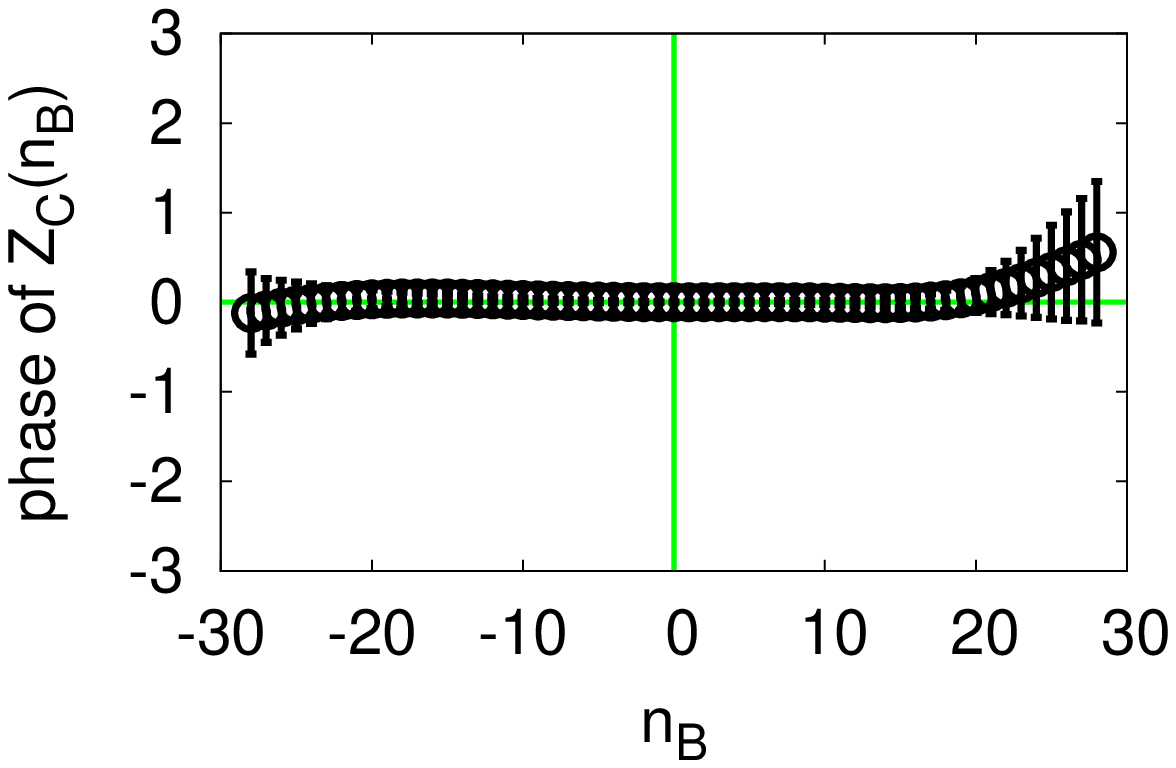}
  \caption{The canonical partition function as a function of the baryon number $n_B$.
Left panel is the absolute value $\log|Z_C(n_B)/Z_C(0)|/(VT^3)$.
From the top $\beta=2.1$ (orange), $1.9$ (red), $1.7$ (magenta), $1.5$
  (green), $1.3$ (dark green), $1.1$ (blue), $0.9$ (black).
Right panel is a phase of $Z_C(n_B)$ at $\beta=1.9$.
}
  \label{fig:Zn}
 \end{center}
\end{figure}

The first numerical result we get is the canonical partition function
$Z_C(n,T,V)$.
We plot $\log|Z_C(n_B,T,V)/Z_C(0,T,V)|/(VT^3)$ and its phase as a
function of the baryon number $n_B$ in Fig.~\ref{fig:Zn}.
The partition function decays very rapidly with $n_B$ and its behavior
changes drastically between $\beta=1.7$ (magenta) and $1.3$
(dark-green), which may correspond to a phase transition.
The phase of the partition function is consistent with zero within the
error bar for $\beta=1.9$ as is shown in the right panel.

\begin{figure}
 \begin{center}
  \includegraphics[width=5.7cm]{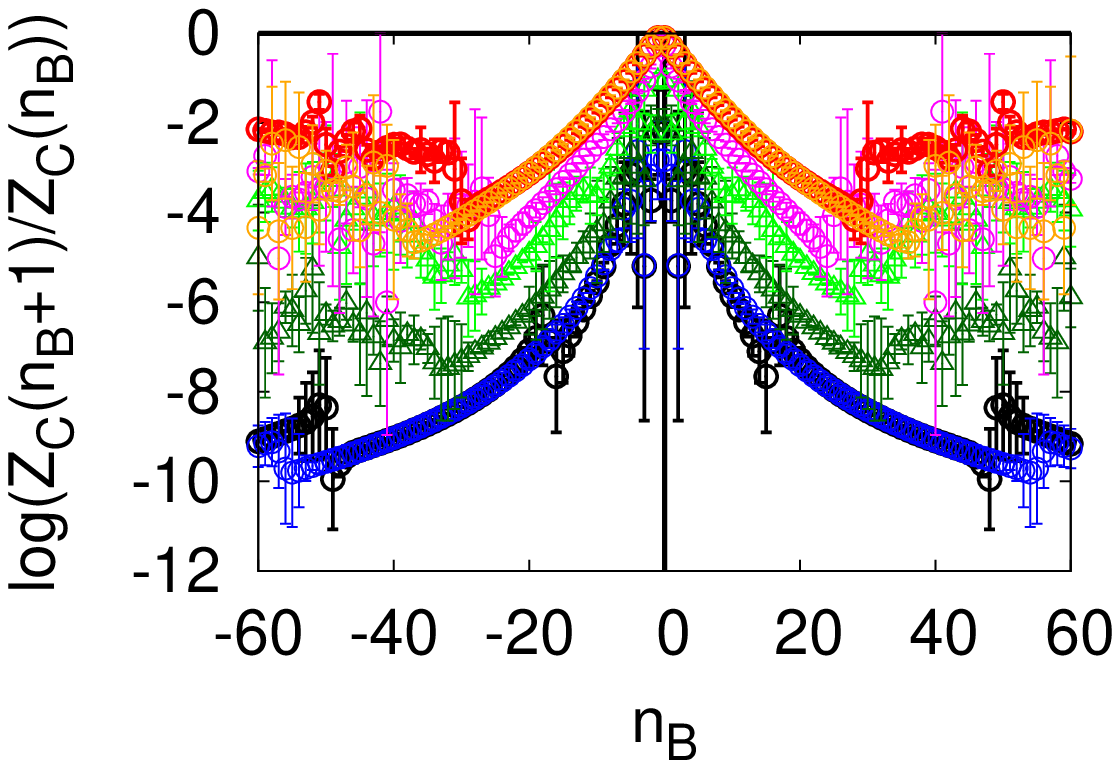}
  \includegraphics[width=5.7cm]{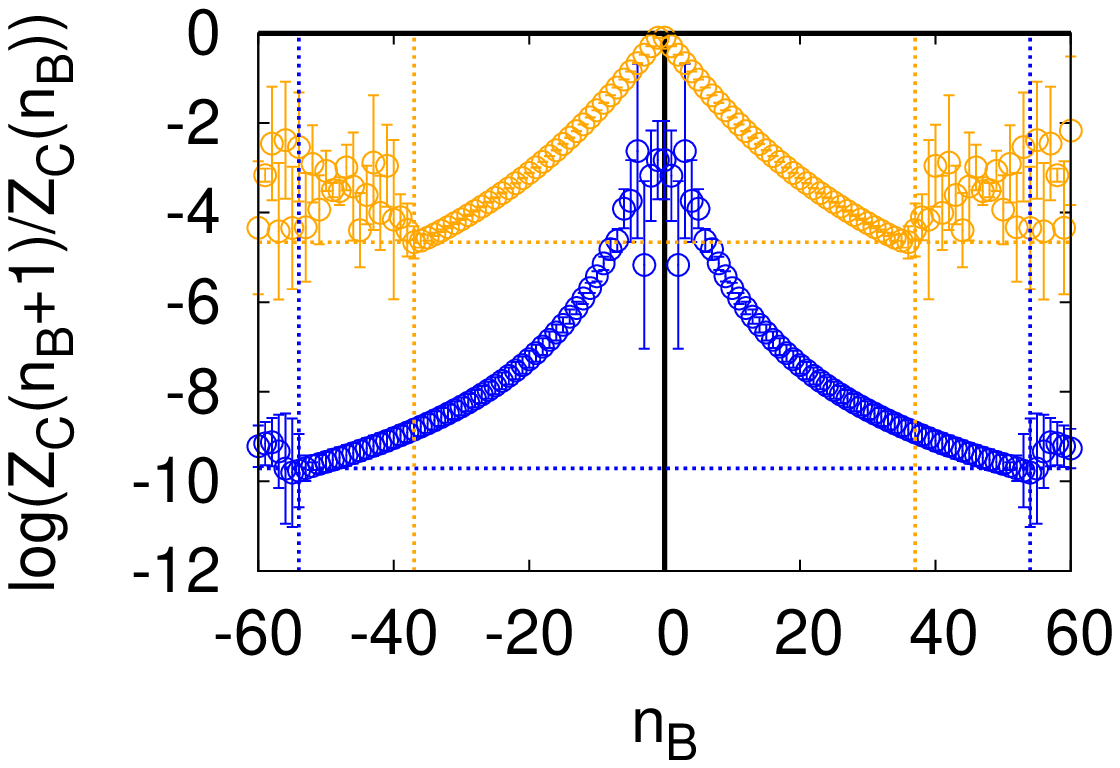}
  \caption{Log plot of $|Z_C(n_B+1)/Z_C(n_B)|$ as a function of the
  baryon number.
  The color and $\beta$ correspondence is the same as in Fig.~2 for left
  panel.
  Right panel is that at $\beta=1.1$ and $2.1$ for example.
}
  \label{fig:dAlambert}
 \end{center}
\end{figure}

The plot range is fixed by using the d'Alembert's convergence condition
\begin{eqnarray}
\lim_{n_B\to\infty}\left|\frac{Z_C(n_B+1)}{Z_C(n_B)}\xi\right|<1,
\end{eqnarray}
\footnote{For negative baryon number we adopt $|Z_C(n_B-1)/Z_C(n_B)|$.}
which gives the convergence radius for the fugacity $\xi$.
The data at $\beta=1.1$ and $2.1$ are plotted in right panel of
Fig.~\ref{fig:dAlambert} for example.
We cut our data at $n_{\rm max}$ where a monotonic decrease stops
indicated by vertical blue and orange lines in the figure.
The horizontal lines show maximal values of the fugacity expected
to be within the convergence radius.
By taking log the line gives our applicable limit for the baryon
chemical potential $\mu_B/T$.
For example we can discuss physics safely at $-10<\mu_B/T<10$ for
$\beta=1.1$ and $-4<\mu_B/T<4$ for $\beta=2.1$ with our method.

\begin{figure}
 \begin{center}
  \includegraphics[width=5.7cm]{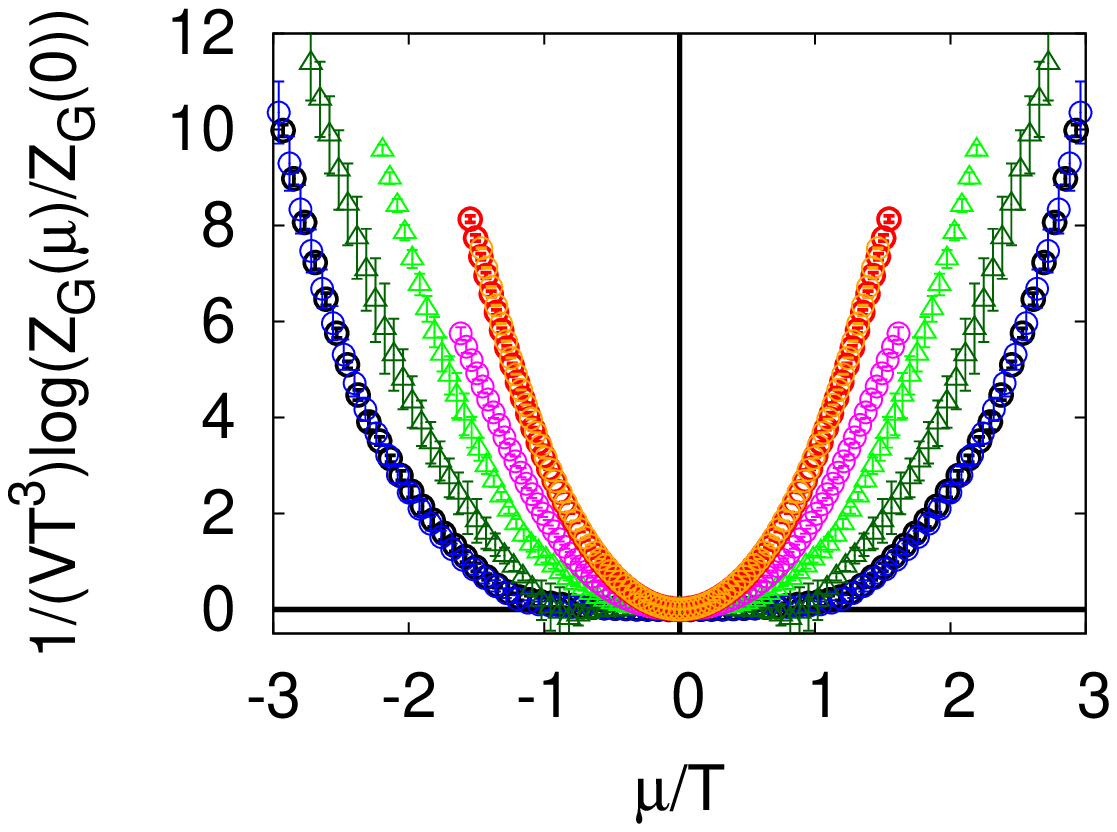}
  \caption{The grand partition function
  $(\log|Z_G(\mu,T,V)/Z_G(0,T,V)|)/V$ as a function of the quark
  chemical potential $\mu/T$.
  The color and $\beta$ correspondence is the same as in Fig.~2.
}
  \label{fig:ZG}
 \end{center}
\end{figure}

The second physical quantity is the grand partition function.
By taking summation for $-n_{\rm max}\le n_B\le n_{\rm max}$ in
(\ref{ZG}) we get the grand partition function for the real chemical
potential.
We plot $(\log|Z_G(\mu,T,V)/Z_G(0,T,V)|)/(VT^3)$ in Fig.~\ref{fig:ZG} as
a function of the quark chemical potential.
According to the statistical physics the logarithm of the grand
partition function divided by the spatial volume is the pressure
\begin{eqnarray}
\frac{P}{T}=\frac{1}{V}\log Z_G.
\end{eqnarray}
The quantity plotted in Fig.~\ref{fig:ZG} is the pressure
$(P(\mu/T)-P(0))/T^4$ in the grand canonical ensemble normalized at
$\mu=0$.

\section{Hadronic observables}

In our procedure the fugacity expansion is based on the hopping parameter expansion.
It may be possible to expand any hadronic operators in terms of the fugacity.
We consider a fugacity expansion of a numerator of some operator VEV
\begin{eqnarray}
&&
\left\langle O\right\rangle_G(\xi,T,V)=\frac{O_G(\xi,T,V)}{Z_G(\xi,T,V)},
\\&&
O_G(\xi,T,V)=\int DU \left\langle O\right\rangle_{\rm quark}(\xi){\rm Det}D_W(\xi;U)e^{-S_G(U)}
=\sum_{n=-\infty}^\infty O_C(n,T,V)\xi^n,
\label{eqn:OG}
\end{eqnarray}
where $\left\langle O\right\rangle_{\rm quark}$ is a Wick contraction of a hadronic operator $O$
in quark fields.
For example we consider the chiral condensate.
It is easy to expand its Wick contraction in terms of the hopping parameter
\begin{eqnarray}
\left\langle\bar\psi\psi\right\rangle_{\rm quark}
=-{\rm tr}\left(\frac{1}{D_W}\right)=-{\rm tr}\left(\frac{1}{1-\kappa Q}\right)
=\sum_{m=0}^\infty\kappa^m{\rm tr} Q^m
=\sum_{n=-\infty}^\infty o_n(U)\xi^n.
\end{eqnarray}
Counting the winding number in temporal direction we get the last equality.
Multiplying the determinant contribution (\ref{regroup}) we apply the
same regrouping procedure in Sec.~\ref{sec:WNE} and get the fugacity
expansion (\ref{eqn:OG}).

\begin{figure}
 \begin{center}
  \includegraphics[width=5.7cm]{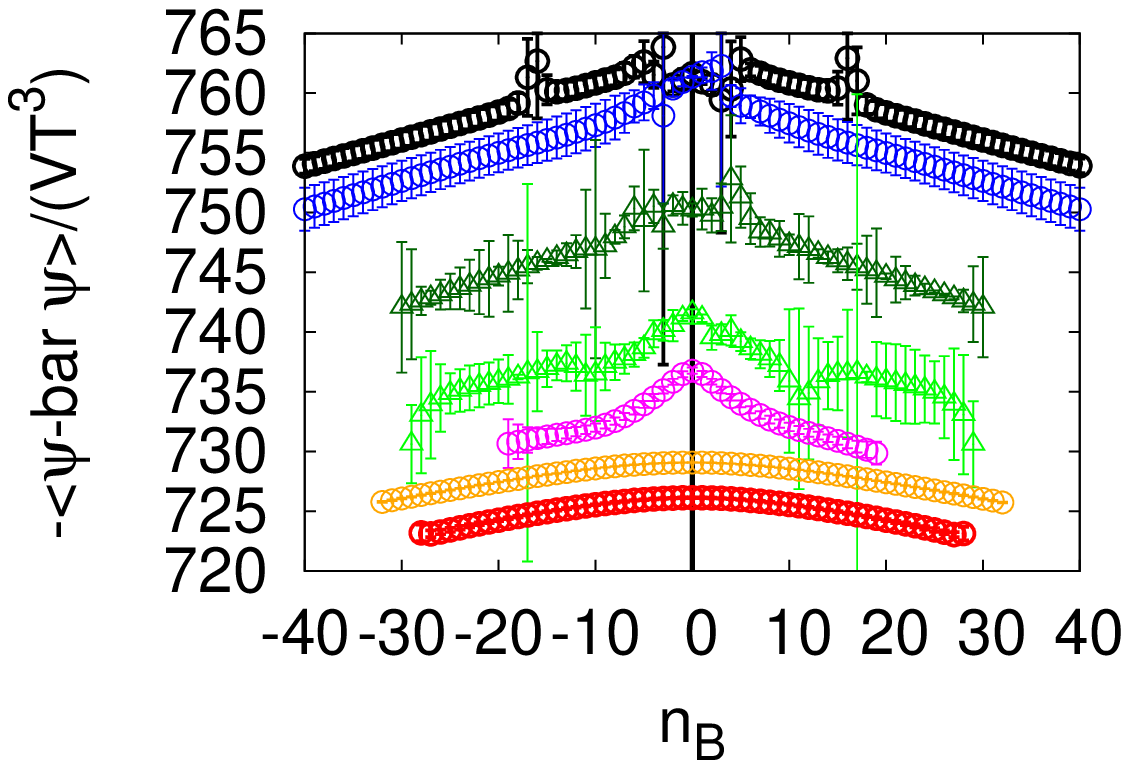}
  \includegraphics[width=5.7cm]{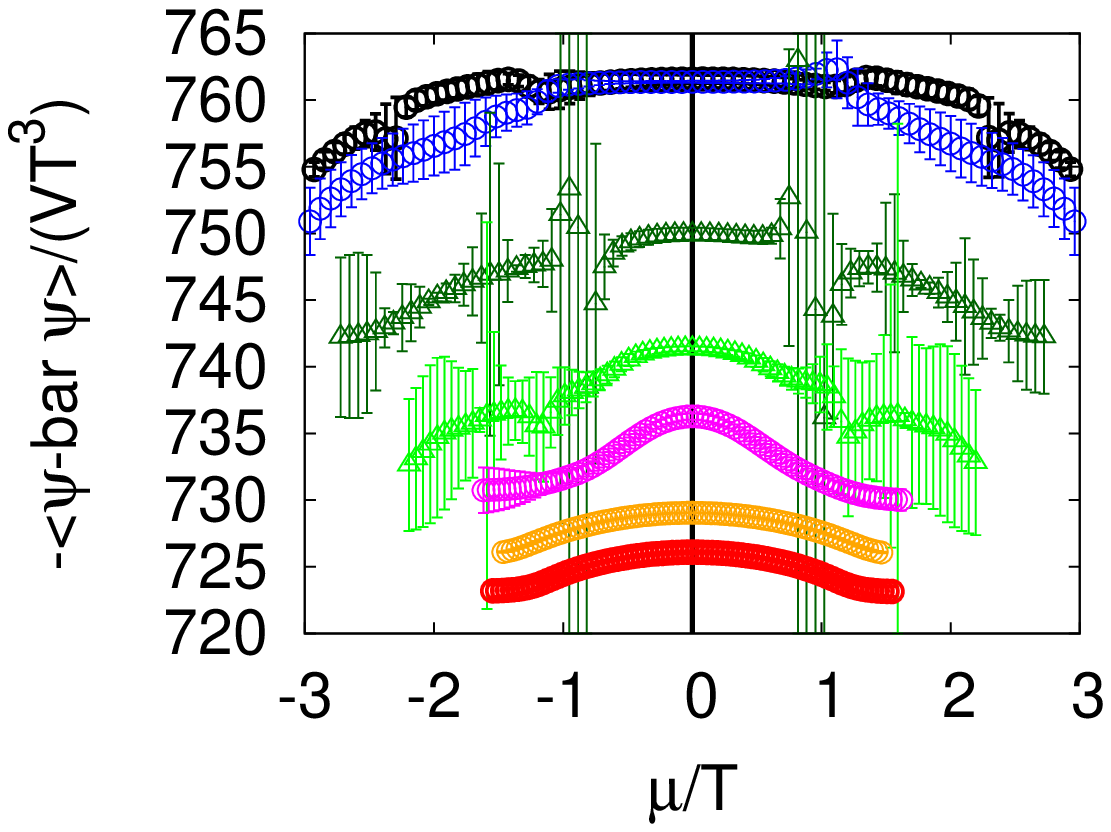}
  \caption{Left panel is the chiral condensate
  $\left\langle\bar\psi\psi\right\rangle/(VT^3))$ in the canonical ensemble as a
  function of the baryon number.
  Right panel is that in the grand canonical ensemble as a function of
  the quark chemical potential $\mu/T$.
The color and $\beta$ correspondence is the same as in Fig.~2.
}
  \label{fig:bpsipsi}
 \end{center}
\end{figure}
Once we have two coefficients $Z_C(n,T,V)$ and $O_C(n,T,V)$ the VEVs of
the operator with the canonical and the grand canonical ensemble are
available.
The canonical ensemble VEV is given by taking the ratio of two
coefficient
\begin{eqnarray}
\left\langle O\right\rangle_C(n,T,V)=\frac{O_C(n,T,V)}{Z_C(n,T,V)}.
\end{eqnarray}
The results for the chiral condensate
$-\left\langle\bar\psi\psi\right\rangle_C/(VT^3)$ are given in the left
panel of Fig.~\ref{fig:bpsipsi} as a function of the baryon number.
A VEV in the grand canonical ensemble is given by taking fugacity
summation with real chemical potential
\begin{eqnarray}
\left\langle O\right\rangle_G(\xi,T,V)
=\frac{\sum_{n=-n_{\rm max}}^{n_{\rm max}}O_C(n,T,V)\xi^n}
{\sum_{n=-n_{\rm max}}^{n_{\rm max}}Z_C(n,T,V)\xi^n}.
\end{eqnarray}
The chiral condensate $-\left\langle\bar\psi\psi\right\rangle_G/(VT^3)$
is given in the right panel of Fig.~\ref{fig:bpsipsi} as a function of
the quark chemical potential $\mu/T$.
The condensate in the figure is a bare quantity without renormalization.
Since we adopted the Wilson fermion we have an additive correction for
$\left\langle\bar\psi\psi\right\rangle$, which is not subtracted in this
paper.

From the right panel of Fig.~\ref{fig:bpsipsi} the chiral restoration
phase transition at finite chemical potential seems to be seen.
A relatively large value around $\mu/T=0$ decreases at large chemical
potential and a fall occurs rapidly at a narrow region.
The would-be transition parameter $\mu/T$ becomes larger for lower
temperature.

\begin{figure}
 \begin{center}
  \includegraphics[width=5.7cm]{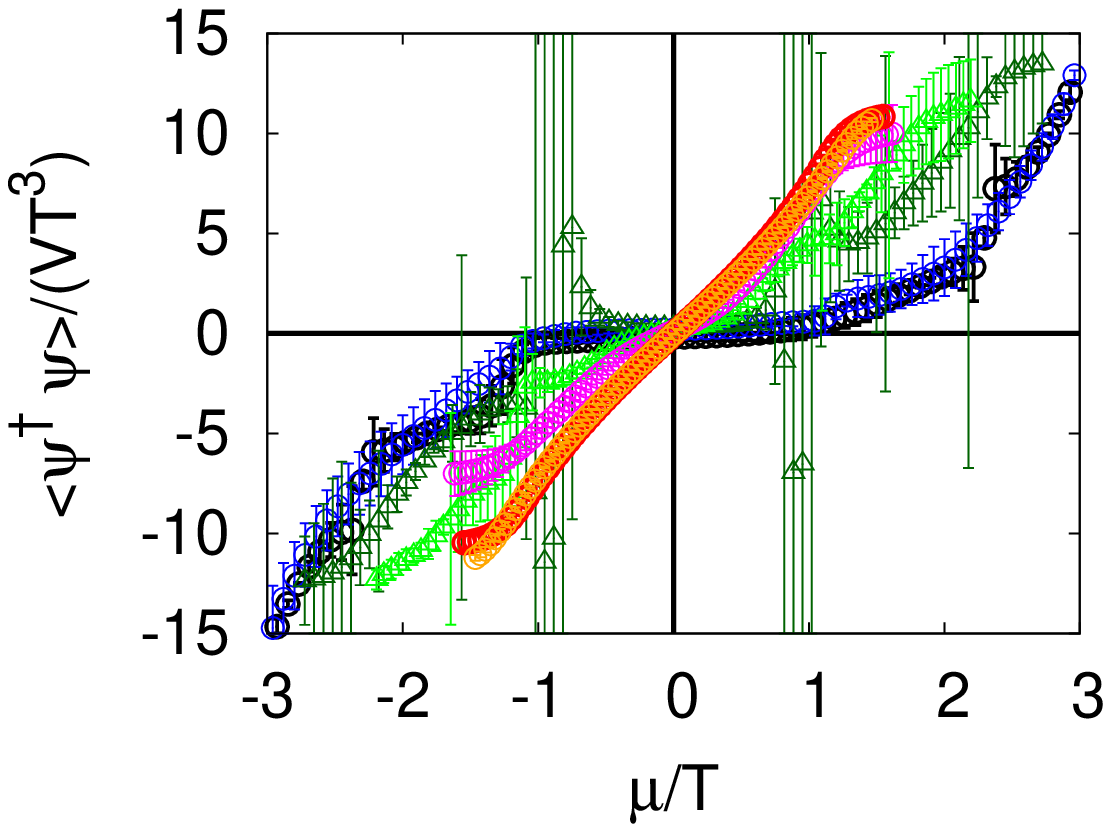}
  \caption{Quark number density
  $-\left\langle\psi^\dagger\psi\right\rangle_G$ as a function of the
  quark chemical potential.}
  \label{fig:nq}
 \end{center}
\end{figure}

In Fig.~\ref{fig:nq} the quark number density
$\left\langle\psi^\dagger\psi\right\rangle_G/(VT^3))$ in the grand
canonical ensemble is plotted as a function of the quark chemical
potential $\mu/T$.

\section{Conclusion}

In this paper we performed the fugacity expansion of the grand partition function by using the hopping
parameter expansion.
This procedure seems to be valid for baryon numbers around $n_B\sim30$
for $8^3\times4$ lattice.
The method is also applied to the numerator of VEVs of hadronic operators.
Taking summation we get a VEV at the real chemical potential.
As an example we evaluate the chiral condensate and show a phase
transition like behavior at high chemical potential for low temperature
region.

This work is done for Zn Collaboration.
This work is supported in part by Grants-in-Aid of the Ministry of
Education (Nos. 26610072. 24340054, 22540265).
This work is in part based on Bridge++ code
(http://suchix.kek.jp/bridge/Lattice-code/).

\end{document}